\documentclass[a4paper,UKenglish,cleveref, autoref, thm-restate]{lipics-v2021}

\title{SAT-IT: an Online Interactive SAT Tracer}

\author{Wilber Bermeo}{Universitat de Girona, Spain}{wilber.bermeo@udg.edu}{https://orcid.org/0009-0001-3748-8164}{}

\author{Jordi Coll}{Universitat de Girona, Spain}{jordicoll@udg.edu}{https://orcid.org/0000-0002-9385-5723}{}
 
\author{Pau Ferrer}{Universitat de Girona, Spain}{u1978930@campus.udg.edu}{https://orcid.org/0009-0001-1244-3402}{}
 
\author{Mateu Villaret}{Universitat de Girona, Spain}{mateu.villaret@udg.edu}{https://orcid.org/0000-0002-8066-3458}{}

\authorrunning{W. Bermeo and J. Coll and P. Ferrer and M. Villaret} 

\Copyright{Wilber Bermeo and Jordi Coll and Pau Ferrer and Mateu Villater} 



\category{Plenary and Shared-Time Demos} 

\relatedversion{} 

\usepackage[normalem]{ulem}
\usepackage{subcaption} 
\usepackage{pdflscape}

\EventEditors{John Q. Open and Joan R. Access}
\EventNoEds{2}
\EventLongTitle{42nd Conference on Very Important Topics (CVIT 2016)}
\EventShortTitle{CVIT 2016}
\EventAcronym{CVIT}
\EventYear{2016}
\EventDate{December 24--27, 2016}
\EventLocation{Little Whinging, United Kingdom}
\EventLogo{}
\SeriesVolume{42}
\ArticleNo{23}

\nolinenumbers
\hideLIPIcs

\begin{document}

\maketitle

\begin{abstract}
Modern Boolean Satisfiability (SAT) solvers, based on the Conflict-Driven Clause Learning (CDCL) paradigm, achieve state-of-the-art efficiency but present a steep learning curve due to their sophisticated algorithms and highly optimized data structures. Understanding these complex mechanics and evaluating the effectiveness of problem encodings is notoriously challenging for students and emerging researchers. To ease this learning process, we introduce the Interactive SAT Tracer (SAT-IT), an open-access web environment designed to make the foundations of SAT solving highly visible and interactive. SAT-IT offers a staged pedagogical progression: from naive backtracking to DPLL and full CDCL with the two-watched literals scheme. Users can clearly inspect fundamental data structures, search space trails, and solving statistics. The tool interactive search space exploration is boosted with literal-level breakpoints for targeted inspection, alongside versatile automatic solving modes that offer both continuous real-time execution and state-based subroutine automation. Combined with a powerful ``what-if'' capability for stepping backward to explore alternative decisions, an instance manager, and an extensible architecture ready to support additional algorithms, SAT-IT serves as a practical, granular lens for experimenting with SAT solving algorithms and analysing encodings efficiency.
\end{abstract}

\section*{Acknowledgments}
This work was funded by MCIN/MICIU/AEI/10.13039/501100011033 and by ERDF A way of making Europe (grants PID2021-122274OB-I00 and PID2024-157625OB-I00).

\section{Introduction}
\label{sec:intro}

Boolean Satisfiability (SAT) has become a powerful and widely used method for tackling difficult combinatorial problems. Its success stems from two key factors: first, a broad range of problem domains can be naturally encoded as SAT instances~\cite{kaufmann2019, kautz1992, demirovic2019, bofill2022constraint}, typically represented as a set of clauses in Conjunctive Normal Form (CNF), and second, the impressive performance achieved by modern SAT solvers in handling these encodings.

This state-of-the-art efficiency is the culmination of decades of research,
resulting in highly sophisticated algorithmic implementations. Modern solvers
rely heavily on Conflict-Driven Clause Learning (CDCL)~\cite{marques2002grasp,
moskewicz2001chaff, marques2009conflict}, augmented by highly optimized data
structures and wisely used inference rules. However, the sophistication 
that makes these algorithms so effective also creates a steep learning curve.
For newcomers, diving into the implementation
details of modern SAT solvers can be a demanding task, which could be eased with a visual support. 

To bridge this gap, we present SAT-IT, an open-access web environment designed
to understand the foundational concepts of CDCL solving and encoding properties. The primary goal of SAT-IT is to make the
mechanics of SAT solving granular and accessible, providing a visual and
intuitive representation of the SAT solving process, explicitly exposing how
fundamental data structures, such as occurrence lists, watch lists, trails, 
are inspected and modified during execution. Furthermore, SAT-IT
exposes a set of relevant statistics of the solving process; such as the number
of visited clauses, propagations, decisions, conflicts, and more. Other relevant
features are discussed below.

The tool offers a staged learning experience by allowing users to interact with
algorithms of increasing sophistication. Users can start with a naive
backtracking algorithm, upgrade to DPLL, and advancing to a vanilla CDCL
procedure with conflict analysis and lemma learning. Additionally, the Two-Watched
Literals (2WL) scheme can be enabled in CDCL.

The platform maintains a detailed trace of the covered search space, listing the branches explored as trails. It provides interactive, step-by-step control over the solving algorithm's execution flow and in subroutines such as the resolution of clauses during the conflict analysis in CDCL or the update of watches when the two-watched literal scheme is enabled. Furthermore, a literal-level breakpoint system allows users to halt the solving procedure precisely when a specific literal is assigned, enabling them to closely inspect the reasons and the consequences of that assignment and gain a deeper understanding of the solver’s behaviour and formula properties.

SAT-IT supports an automated solving procedure with two distinct modes. The first provides a continuous low-level step-by-step execution, where the UI is updated in real-time to reflect each phase of the solving process. The second is an integrated interactive mode, where subroutines execution involve the call of many step-by-step automatically until a defined state is reached, and then, the UI is updated. This allows for accelerated navigation through the search space by bypassing repetitive low-level operations while maintaining granular control.

The tool supports non-linear exploration---as if it was a ``\textit{what if} analysis''---by allowing users to revert the solver until a chosen decision. This is implemented by re-simulating the preceding assignments---decisions and their corresponding propagations. Once re-simulated, users can manually select an alternative branching literal to observe how different would have influenced the search trajectory.

SAT-IT supports the upload of DIMACS CNF formulas, their visualization within the interface, and their removal when no longer needed. Additionally, the system includes a collection of preloaded instances available by default.

The remainder of this document is organised as follows. Section~\ref{sec:preliminaries} present SAT preliminaries for the correct reading of the paper. Section~\ref{sec:algortihms} provides a summary of the SAT solving algorithms that the tool implements. Section~\ref{sec:architecture} explains the architecture that makes the SAT-IT tool unique. In Section~\ref{sec:functionalities} the main visual components are presented. Section~\ref{sec:flow-management} presents a SAT-IT workflow example. Lastly, in Section~\ref{sec:demonstration} we discuss the topics of interest that are planned to be presented during the Plenary and Shared-Time Demos presentation.

The SAT-IT source code is hosted on GitHub\footnote{\url{https://github.com/udg-lai/sat-it}} and is publicly accessible at the project’s domain\footnote{\url{https://udg-lai.github.io/sat-it/}}.

\section{Preliminaries}
\label{sec:preliminaries}
The Boolean Satisfiability problem (SAT) is the paradigmatic NP-complete problem. It consists of deciding the existence of an assignment of truth values to Boolean variables that evaluates a given Boolean formula to true (satisfies the formula). The given formula is usually assumed to be in Conjunctive Normal Form (CNF).

A CNF is defined as a set of constraints over a set of Boolean variables ($x_1,x_2,...,x_n$) that can take a boolean value, i.e. true or false. Each one of these constraints is expressed as a clause, i.e., a disjunction between literals. A literal can be the positive representation of a variable ($x_1$) or its negation ($\lnot x_1$)---alternatively, a negated literal is expressed as $\overline{x}_1$. 

For a CNF to be satisfied by an assignment, each of its clauses must be satisfied, and for a clause to be satisfied, at least one of its literals must be satisfied. A positive literal is satisfied if its variable takes value true in the assignment whilst a negative literal is satisfied if its variable takes value false. If an assignment satisfies a CNF, we call it a model.

For instance, given the CNF with three clauses: $F = \{x_1 \lor \lnot x_2, x_1 \lor x_3, \lnot x_1 \lor \lnot x_2 \lor \lnot x_3\}$ where $x_1,x_2,x_3$ are Boolean variables, $\lor$ is the disjunction operator, and $\lnot$ is the negation, the assignment $\{x_1 = true, x_2 = true, x_3 = false\}$ satisfies $F$ but the assignment $\{x_1 = true, x_2 = true, x_3 = false\}$ does not.

\section{SAT Solving Algorithms}
\label{sec:algortihms}

In this section, we present the pseudocode of the algorithms that SAT-IT includes.

\subsection{Backtracking}

The Backtracking algorithm incrementally constructs partial candidates for a model and abandons a candidate (i.e., backtracks) as soon as it determines that the candidate cannot possibly lead to a valid solution, because of the candidate falsifying a clause. When backtracking occurs, the algorithm returns to the most recent decision point and explores alternative paths.

Backtracking is considered a complete algorithm because it systematically explores the entire search space for possible solutions. Recursive and iterative descriptions of backtracking are widely known. In Figure~\ref{fig:bkt-pc} we present the iterative version of Backtracking.

\begin{figure}[h]

\scriptsize
    \begin{lstlisting}[escapeinside={(*}{*)},tabsize=2,frame=single]
    (*\textbf{ALGORITHM}*) Backtracking(F : CNF) (*\textbf{RETURNS}*) (SAT/UNSAT, MODEL)
    	trail := []
    	(*\textbf{WHILE}*) nonAssignedVars(trail, F) (*\textbf{OR}*) falsifiedClause(trail, F)
    		(*\textbf{IF}*) falsifiedClause(trail, F)
    			(*\textbf{IF}*) noDecisions(trail)
    				(*\textbf{RETURN}*) (UNSAT, _)
    			(*\textbf{ELSE}*)
    				l := lastLiteralDecided(trail)
    				dl := decisionLevel(l, trail)
    				backtrack(trail, dl) 
    				assign(trail, (*$\neg \text{l}$*))
    			(*\textbf{EIF}*)
    		(*\textbf{ELSE}*)
    			l := chooseNonAssignedLiteral(trail, F)
    			decide(trail, l) 
    		(*\textbf{EIF}*)
    	(*\textbf{EWHILE}*)
    	(*\textbf{RETURN}*) (SAT, trail)
    (*\textbf{EALGORITHM}*)
    \end{lstlisting}
    \caption{Backtracking pseudocode.}
    \label{fig:bkt-pc}    
\end{figure}

The set of partial candidates (in SAT, the branching variables) can be represented as nodes in a search tree. The algorithm performs a depth-first traversal of this tree, beginning at the root. At each node \(n\), it verifies whether the current partial solution is still valid. If it is not, the algorithm backtracks and prunes that branch. If the node represents a complete and valid solution, a model has been found.

Although backtracking can be used to solve SAT problems, it is highly inefficient on its own. Many branches of the search tree can be predicted to lead to failure, yet pure backtracking still explores them, resulting in considerable wasted effort.

\subsection{DPLL}

\begin{figure}
\scriptsize
    \begin{lstlisting}[escapeinside={(*}{*)},tabsize=2,frame=single]
    (*\textbf{ALGORITHM}*) DPLL(F : CNF) (*\textbf{RETURNS}*) (SAT/UNSAT, MODEL)
    	trail := []
        (*\textcolor{blue}{UnitPropagation(trail, F)}*)
    	(*\textbf{WHILE}*) nonAssignedVars(trail, F) (*\textbf{OR}*) falsifiedClause(trail, F)
    		(*\textbf{IF}*) falsifiedClause(trail, F)
    			(*\textbf{IF}*) noDecisions(trail)
    				(*\textbf{RETURN}*) (UNSAT, _)
    			(*\textbf{ELSE}*) 
    				l := lastLiteralDecided(trail)
    				dl := decisionLevel(l, trail)
    				backtrack(trail, dl) 
    				assign(trail, (*$\neg \text{l}$*)) 
    			(*\textbf{EIF}*)
    		(*\textbf{ELSE}*)
    			l := chooseNonAssignedLiteral(trail, F)
    			decide(trail, l)
    		(*\textbf{EIF}*)
            (*\textcolor{blue}{UnitPropagation(trail, F)}*)
    	(*\textbf{EWHILE}*)
    	(*\textbf{RETURN}*) (SAT, trail)
    (*\textbf{EALGORITHM}*)
    \end{lstlisting}
    \caption{DPLL pseudocode.}
    \label{fig:dpll-pc}
\end{figure}

The Davis–Putnam–Logemann–Loveland (DPLL) algorithm introduced in~\cite{nieuwenhuis2006solving} is a complete backtracking-based algorithm used to decide the satisfiability of formulas in CNF.

DPLL extends basic Backtracking by integrating \emph{unit propagation}(UP). If during the search for a satisfying interpretation, a clause becomes unit---that is, it contains exactly one unassigned literal while all others are assigned false---then the unassigned literal must be assigned true to satisfy the clause. This inference step significantly prunes the search tree without sacrificing completeness.

In many cases, a single unit clause can trigger a cascade of further unit propagations, drastically reducing the number of explicit decisions required during the search.

As with Backtracking, DPLL can be implemented both recursively and iteratively. The iterative version is shown in Figure~\ref{fig:dpll-pc}.

The key difference from standard Backtracking is highlighted in blue: the function \texttt{Unit Propagation}, which performs a unit propagation---if possible---after every assignment that did not lead to a conflict.

Despite its improvements over simple Backtracking, DPLL has a notable limitation: it does not retain information about previous conflicts. As a result, the algorithm may do assignments that will led to already visited conflicts.

\subsection{CDCL}

\begin{figure}[t]
\footnotesize
    \begin{lstlisting}[escapeinside={(*}{*)},tabsize=2,frame=single]
    (*\textbf{ALGORITHM}*) CDCL(F : CNF) (*\textbf{RETURNS}*) (SAT/UNSAT, MODEL)
    	trail := []
        (*\textcolor{blue}{UnitPropagation(trail, F)}*)
    	(*\textbf{WHILE}*) nonAssignedVars(trail, F) (*\textbf{OR}*) falsifiedClause(trail, F)
    		(*\textbf{IF}*) falsifiedClause(trail, F)
    			(*\textbf{IF}*) noDecisions(trail)
    				(*\textbf{RETURN}*) (UNSAT, _)
    			(*\textbf{ELSE}*) // Learn and backjump
    				(*\textcolor{blue}{cl := falsifiedClauseObtention(trail, F)}*)
    				(*\textcolor{blue}{(dl, lemma) := conflictAnalysis(trail, F, cl)}*)
    				(*\textcolor{blue}{learn(lemma, F)}*)
    				(*\textcolor{blue}{backjump(trail, dl)}*)
    			(*\textbf{ENDIF}*)
    		(*\textbf{ELSE}*)
    			l := chooseNonAssignedLiteral(trail, F)
    			decide(trail, l)
    		(*\textbf{ENDIF}*)
            (*\textcolor{blue}{UnitPropagation(trail, F)}*)
    	(*\textbf{ENDWHILE}*)
    	(*\textbf{RETURN}*) (SAT, trail)
    (*\textbf{ENDALGORITHM}*)
    \end{lstlisting}
    \caption{CDCL pseudocode.}
    \label{fig:cdcl-pc}
\end{figure}

CDCL algorithm (see Figure~\ref{fig:cdcl-pc}) sophisticates the DPLL framework incorporating two powerful techniques: \textit{clause learning} and \textit{backjumping}.

Although these additions increase its complexity, CDCL remains a complete algorithm and significantly improves the performance through conflict-driven pruning.

When a conflict is detected, CDCL performs \textit{conflict analysis} to determine the causes of the inconsistency. Using the \textit{resolution rule}, it derives a new clause---known as a \textit{learned clause} or \textit{lemma}---that logically follows from the original formula. This clause is then added to the formula, effectively preventing the solver from trying the same conflictive assignment.

Unlike DPLL, which always performs chronological backtracking, CDCL uses the learned clause to identify the most relevant decision level to return to, enabling \textit{non-chronological backtracking} or backjumping. This allows the solver to skip irrelevant parts of the search tree and focus on unexplored paths.

\section{Architecture}\label{sec:architecture}

In this section, we provide some intuitions of the software architecture upon which SAT-IT. 

\begin{figure}[t]
    \centering
    \resizebox{0.8\linewidth}{!}{%
    \centering
    \includegraphics[width=\linewidth]{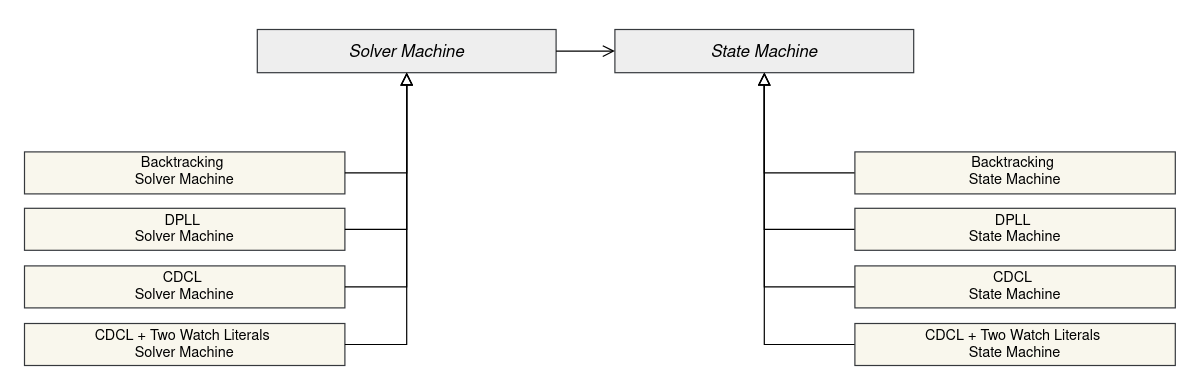}
    }
    \caption{Solver Machine and State Machine Composition.}
    \label{fig:solver_machine}
\end{figure}

SAT-IT is a pedagogical tool that provides deep, granular insight into the internal mechanics of the implemented SAT solving algorithms by showing how data structures are updated trough the solving process. Moreover, SAT-IT has been thought with extensibility in mind, enabling future algorithms and paradigms to be included in the application.

To support robust user–tool interaction—including solver-generated events,
atomic state inspection, and dynamic execution control—the system requires a
clear separation of concerns and a high degree of decoupling between logical components.
To achieve this, each SAT solving procedure in SAT-IT is modelled as a state
diagram (or state machine)\footnote{A deeper explanation of the architecture and the state diagrams corresponding to the solving algorithms considered in the tool can be found in the Appendix.}, extending
the rule-based framework presented in~\cite{satitjava}. In SAT-IT, these
rule-based transitions are further augmented to support user-triggered events
that may alter the execution flow. 

Conceptually, a solver in SAT-IT is an event-driven system that is constructed upon a
\emph{solver machine} which encapsulates a specific \emph{state machine} (see Figure~\ref{fig:solver_machine}). The
state machine defines the logical states and transition rules for a given
algorithm; for instance, the CDCL state machine includes states for lemma
learning and non-chronological backjumping that are absent in DPLL or
Backtracking machines, though they share common operational states such as the
decision phase. The solver machine acts as a controller, navigating the
transitions defined by the state machine, advancing the computation, emitting
events to the user interface, and reacting to user-triggered actions. All solver
machines derive from a common abstract solver machine, providing a uniform API
for state transitions, inspection, and event handling. This abstraction ensures
system-wide consistency and facilitates the integration of new solving
techniques. 

\section{Functionalities}
\label{sec:functionalities}

In this section, we present a detailed explanation of the most important
visualization mechanisms and their data structures for the three algorithms
included in SAT-IT

\subsection{The Trail}
\label{subsec:trail-design}

SAT-IT adopts the operational framework originally introduced
in~\cite{nieuwenhuis2006solving} and subsequently adapted in~\cite{satitjava},
which introduces the rule-based transition system that we use as reference for
formal representation of the trail. Within this framework, a trail is a
sequence of literals that represents the literals satisfied by an
interpretation, ordered chronologically by assignment. This rule-based transition system enables
formal reasoning about Backtracking, DPLL, and CDCL algorithms through a concise
operational notation. 

Within the adopted framework, the trail is transformed through the application
of four rules: \emph{Decision}, \emph{Unit Propagation}, \emph{Backtrack}, and
\emph{Backjump}. The \emph{Decision} rule assigns a truth value to an unassigned
literal; by default, SAT-IT assigns variables in ascending order and \emph{true}
value. The \emph{Unit Propagation} rule enforces the assignment of literals
implied by unit clauses---those in which all but one literal have been falsified
and remains unassigned. When a conflict is found, the \emph{Backtrack} rule
reverts the most recent decision. The \emph{Backjump} rule is exclusive of CDCL,
and is invoked after the \emph{Learn} rule. Although multiple decision levels (a
decision level represents the number of previous decisions made) may exist where
the learned clause becomes a unit, the second highest decision level present
within the clause is targeted. This ensures that the learned clause becomes unit
at the earliest possible stage. Notice that the \emph{Decision} and the
\emph{Unit Propagation} rules extend the trail and the \emph{Backjump} and
\emph{Backtrack} rules may shrink it. For an exhaustive discussion of this
formal framework, the reader is referred to~\cite{satitjava}.

\begin{figure}[t]
    \centering
    \includegraphics[width=0.5\linewidth]{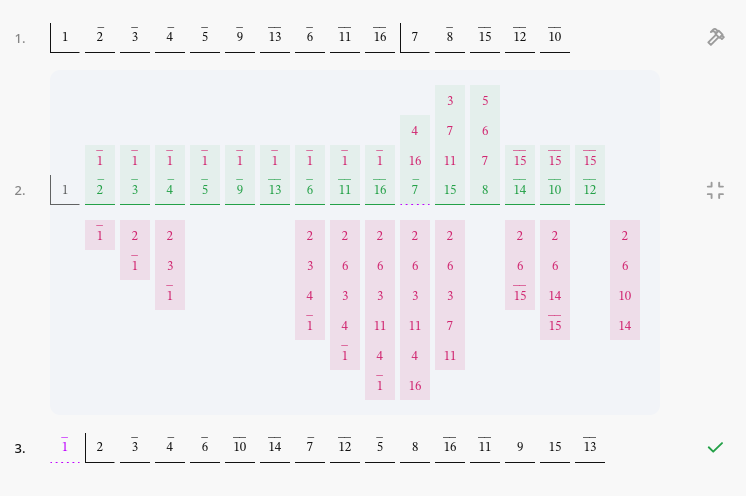}
    \caption{Search space by means of trails. Focusing on trail number two, on the top of it is possible to appreciate the reasons that propagated the literals, whereas under it the conflictive analysis, with the conflictive, derived, and learned clauses are displayed from right to left.}
    \label{fig:trails}
\end{figure}

In SAT-IT, a trail (see Figure~\ref{fig:trails}) is represented as a sequence of literals categorized
by their assignment type, i.e., the rule used to assign them.
A decision is denoted by a box-shaped boundary that signifies the start of a new
\textit{decision level}. Decisions might be displayed within a
\textit{collapsed} box (represented by two vertical lines and a bottom line),
which hides subsequent propagations within that decision level, or
by and \textit{expanded} box (represented by a left vertical line and a bottom
line), exposing all UPs associated within that level. These visual
states for decided literals are interactive, allowing the user to toggle between
collapsed and expanded views. Propagated literals are distinguished solely by a
solid bottom line. A distinct case is also illustrated, at the beginning of the trail number three, a purple-coloured literal appear with
dashed bottom lines. This visual style represents a backtracking in the context
of DPLL or Backtracking algorithms. In CDCL, a literal of this kind represents the complementary of the First
Unique Implication Point (1UIP) identified during \emph{conflict analysis} in
the previous trail, and hence the literal that has been propagated after backjumping.

As shown in  Figure~\ref{fig:trails}, in trail two, a SAT-IT trail is accompanied by
two more views designed to provide deeper insight of the solving procedure. The
view situated on the top and spatially united to the trail shows the
\emph{antecedents}---or reasons---of propagated literals, where each column is a
clause. In case the current trail does not satisfy the formula, the bottom view
shows the \emph{conflictive clause} at the right of the trail. In CDCL, at the
left of the conflictive clause, SAT-IT shows the clauses derived during the
conflict analysis. More precisely, from right to left, it shows below each
propagation reason the clause resulting from applying a resolution step between
the reason and the previously derived clause at the right. Note that the
left-most derived clause is the lemma that will be learned, and it is displayed
below the 1UIP. In both views, clauses (reasons, conflictive clause, resolvent
clauses, and the derived lemma), are represented as a vertical coloured box,
green (satisfied) or red (unsatisfied) depending on the formula interpretation.
The colour scheme is applied consistently also to literals in the clauses. Also,
both views are updated incrementally in real time as new propagated literals are
added to the trail (top view) or resolution steps are applied during conflict
analysis (bottom view). A trail is accompanied by a
clickable button on its right to show or hide the reasons and the conflict
analysis steps.

SAT-IT as a visual tracer requires an effective representation of the search
space. To this end, we represent the solving process of an instance $F$ as an
enumerated sequence of trails $(M_1, \dots, M_n)$ as it can be seen in
Figure~\ref{fig:trails} with the number located at the left-side of each trail. In particular, when the solving process
is completed,  $M_i \not\models F$ for all $i < n$, and if $F$ is satisfiable $M_n$
is a model. 

\subsection{Occurrence and Watch Lists}
\label{sec:occurrence-watch}

\begin{figure}[t]
    \centering
    \begin{subfigure}[b]{0.3\linewidth}
        \centering
            \includegraphics[width=0.5\linewidth]{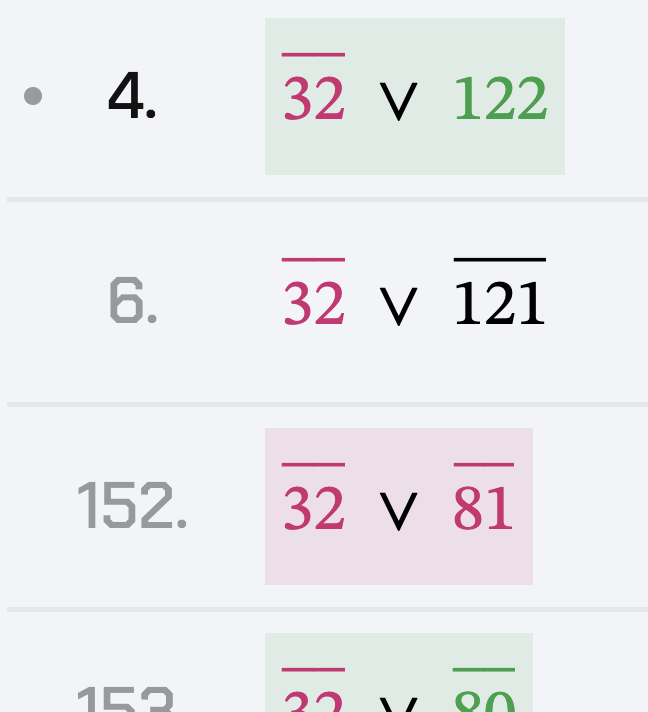}
        \label{fig:occurrence-list}
    \end{subfigure}
    \hspace{1cm} 
    \begin{subfigure}[b]{0.3\linewidth}
        \centering
        \includegraphics[width=0.7\linewidth]{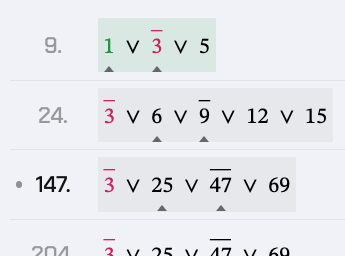}
        \label{fig:watch-list}
    \end{subfigure}
    
    \caption{Visualization of data structures. Left) Occurrence list of literal $\overline{32}$. Clause $152$ is currently unvisited yet already falsified. Right) Watch list of literal $\overline{3}$. Triangle indicators denote watched literals. As clauses are visited the watches may change.}
    
    \label{fig:list}
\end{figure}

The standard mechanism in modern solvers to apply UP while
checking if a conflict has been reached during the search is to inspect the new
literals in the trail in a FIFO basis. For this purpose, a pointer or
\emph{head} specifies which literal is inspected, for example, in
Figure~\ref{fig:list} (Left), the head points to literal $32$. The pointer is
moved to the right after a literal has been inspected.  The inspection of a
literal $l$ consists in visiting the clauses containing its complement
$\overline{l}$, since these are the clauses that might have become unit or
unsatisfied after falsifying $\overline{l}$. For this purpose, SAT-IT lets the
user configure the algorithm to use either watch lists (the standard and most
efficient data structure) or full occurrence lists, thus enabling a comparative
of performances and providing a visual motivation for the use of 2WL scheme.

The occurrence list is a visual component situated in the tool pane (see
Figure~\ref{fig:list}). SAT-IT facilitates interactive navigation through this
list with a pointer that indicates which clause is currently being visited. The
satisfaction or unsatisfaction of the clauses in the list is visually
represented  employing the red/green colour-coding scheme previously established.
To provide further clarity during inspection, visited clauses that are not
already satisfied or unsatisfied are highlighted with a gray background.
Moreover, when UP occurs due to the currently visited clause, the
colour scheme of all affected clauses is updated in real time.

If 2WL scheme is enabled, instead of the occurrence list, the watch list is
displayed, showing only those clauses where the complementary literal is
actively being watched as shown in Figure~\ref{fig:list} (Right). In this watch list
view, specific indicators are rendered below the 2WL, which are dynamically
updated as clauses are inspected and watches are updated because of the propagation
procedure. Furthermore, in this specific case, the tool provides a dedicated
visualization area for clauses bypassed by 2WL, offering user acknowledgment of
the efficiency and behaviour of the watched literal data structures.

\section{Flow Management}
\label{sec:flow-management}
In this section, we highlight SAT-IT's workflow. To this end we
illustrate it by exploring a hypothetical execution in CDCL as it extends the other algorithms of
the tool.

\begin{figure}[h]
    \centering
    \begin{subfigure}[b]{0.22\linewidth}
        \centering
        \includegraphics[width=0.34\linewidth]{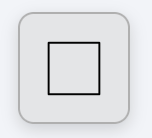}
        \caption{Preprocessing.}
        \label{fig:preprocessing}
    \end{subfigure}
    \hfill
    \begin{subfigure}[b]{0.22\linewidth}
        \centering
        \includegraphics[width=0.6\linewidth]{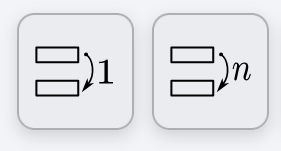}
        \caption{Clause list options.}
        \label{fig:cd-basic}
    \end{subfigure}
    \hfill
    \begin{subfigure}[b]{0.22\linewidth}
        \centering
        \includegraphics[width=0.6\linewidth]{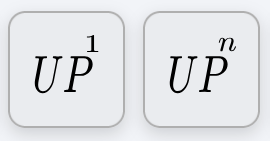}
        \caption{Unit propagation.}
        \label{fig:cd}
    \end{subfigure}
    \hfill
    \begin{subfigure}[b]{0.22\linewidth}
        \centering
        \includegraphics[width=0.7\linewidth]{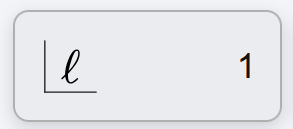}
        \caption{Decision.}
        \label{fig:decision}
    \end{subfigure}

    \vspace{0.5cm}

    \begin{subfigure}[b]{0.22\linewidth}
        \centering
        \includegraphics[width=0.52\linewidth]{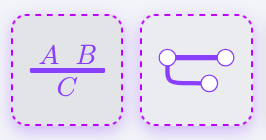}
        \caption{Conflict analysis.}
        \label{fig:ca}
    \end{subfigure}
    \hfill
    \begin{subfigure}[b]{0.22\linewidth}
        \centering
        \includegraphics[width=0.8\linewidth]{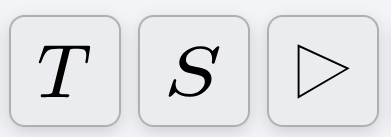}
        \caption{Automatic options.}
        \label{fig:auto}
    \end{subfigure}
    \hfill
    \begin{subfigure}[b]{0.22\linewidth}
        \centering
        \includegraphics[width=\linewidth]{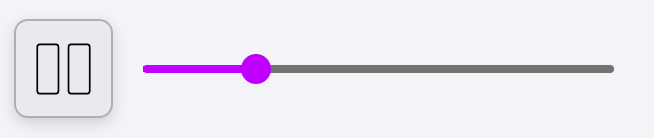}
        \caption{Automatic mode.}
        \label{fig:auto-step-by-step}
    \end{subfigure}
    \hfill
    \begin{subfigure}[b]{0.22\linewidth}
        \centering
    \includegraphics[width=0.8\linewidth]{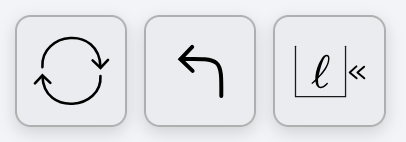}
        \caption{Non-algorithmic.}
        \label{fig:misc}
    \end{subfigure}

    \caption{Relevant interactive buttons.}
    \label{fig:buttons}
\end{figure}

After selecting a problem, SAT-IT will activate the preprocessing phase,
where a single option is available to identify all empty and unary clauses (clauses with no literals and with just one literal, respectively) (see
Figure~\ref{fig:preprocessing}) Note that this process may already determine that the formula is unsatisfiable. If this is not the case, every unary clause will add a literal to the trail by UP, and the solving process will start: the head will point to the first literal $l$ in the trail, and the solver will transition into the \textit{Conflict Detection} phase. This phase is in charge of inspecting the list of clauses containing $\overline{l}$ (either the occurrence list or the watched list), as detailed in Section~\ref{sec:occurrence-watch}, to find unit clauses (propagating a literal) or falsified clauses (finding a conflict).
At this stage, the SAT-IT interface offers
the user several inspection options in which visiting a single or all clauses from
the clause list is possible (see Figure~\ref{fig:cd-basic}).

Upon completing the clause list inspection, if no conflicts were detected, the head of the tail will point to the next literal, if any exists. To speed up the process, the SAT-IT interface offers the user the possibility to either find the next UP or execute all possible UPs (see Figure~\ref{fig:cd}).

If no conflictive clauses were detected during the \textit{Conflict Detection}
phase, the state machine transitions into the \textit{Decision} phase. By
pressing the decision button (see Figure~\ref{fig:decision}) the literal
displayed at the right of it will be decided and, as a consequence, the
solver will enter the \textit{Conflict Detection} phase. If a conflict is identified, the solver transitions to the \textit{Conflict Analysis} stage. Within this stage, the user can either execute a single resolution step or automatically find and propagate the 1UIP to resolve the conflict (see Figure~\ref{fig:ca}). The recursive application of the Decision, Conflict Detection and Conflict Analysis phases, constitute the CDCL algorithm.

SAT-IT also offers the option to explore ``what-if'' scenarios by undoing the execution of the algorithm until a previous decision point (removing all trails after this point, and cancelling all assignments after this decision). This can be done by clicking on a decided literal and selecting the ``Revert Up to Here'' option from the displayed pop-up, or by pressing the \emph{undo} button (see Figure~\ref{fig:misc}). After undoing the decision, the user can explore another search path by replacing the decision placeholder with an alternative literal. 

Furthermore, SAT-IT provides an \emph{automatic solving} functionality. This functionality might be used to finish the current trail, or to solve the entire instance and display the full searched space as a sequence of trails. Both things can be emulated by the automated \emph{step-by-step} mode, where users can observe assignments, visited clauses, statistics, and other relevant information as they occur. The execution speed can be adjusted, and the solver can be halted at any time (see Figures~\ref{fig:auto} and~\ref{fig:auto-step-by-step}).

\section{Demonstration}
\label{sec:demonstration}

In this section, we present the content that will be transmitted in workshop.

The session will begin with a brief presentation of SAT solving preliminaries to ensure a common baseline for the audience. 
To illustrate the limitations of naive search, we will walk through a backtracking algorithm using a CNF instance easy to solve. 
This example will serve as our framework for introducing essential components and concepts like trails, 
decision levels, and conflicts. For instance, the CNF formula could be:

$$F = (3 \lor 4 \lor \overline{1} \lor 5) \land (\overline{3} \lor 4 \lor 5) \land (3 \lor \overline{4} \lor 1) \land (1 \lor 2) \land (1 \lor \overline{2}) \land (\overline{1} \lor \overline{5}) \land (\overline{3} \lor \overline{4} \lor 5)$$

\begin{figure}[t]
    \centering
    \begin{minipage}{\linewidth}
        \centering
        
        \begin{minipage}{\linewidth}
            \centering
            \begin{subfigure}{0.48\linewidth}
                \centering
                \includegraphics[width=\linewidth]{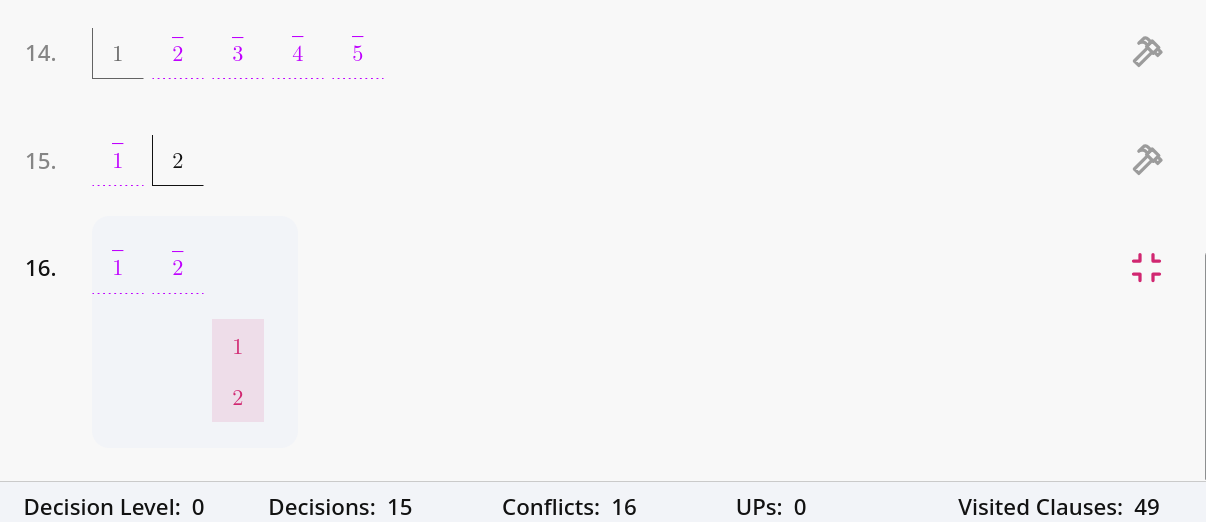}
                \caption{Backtracking}
                \label{fig:dpll-a}
            \end{subfigure}
            \hfill
            \begin{subfigure}{0.48\linewidth}
                \centering
                \includegraphics[width=\linewidth]{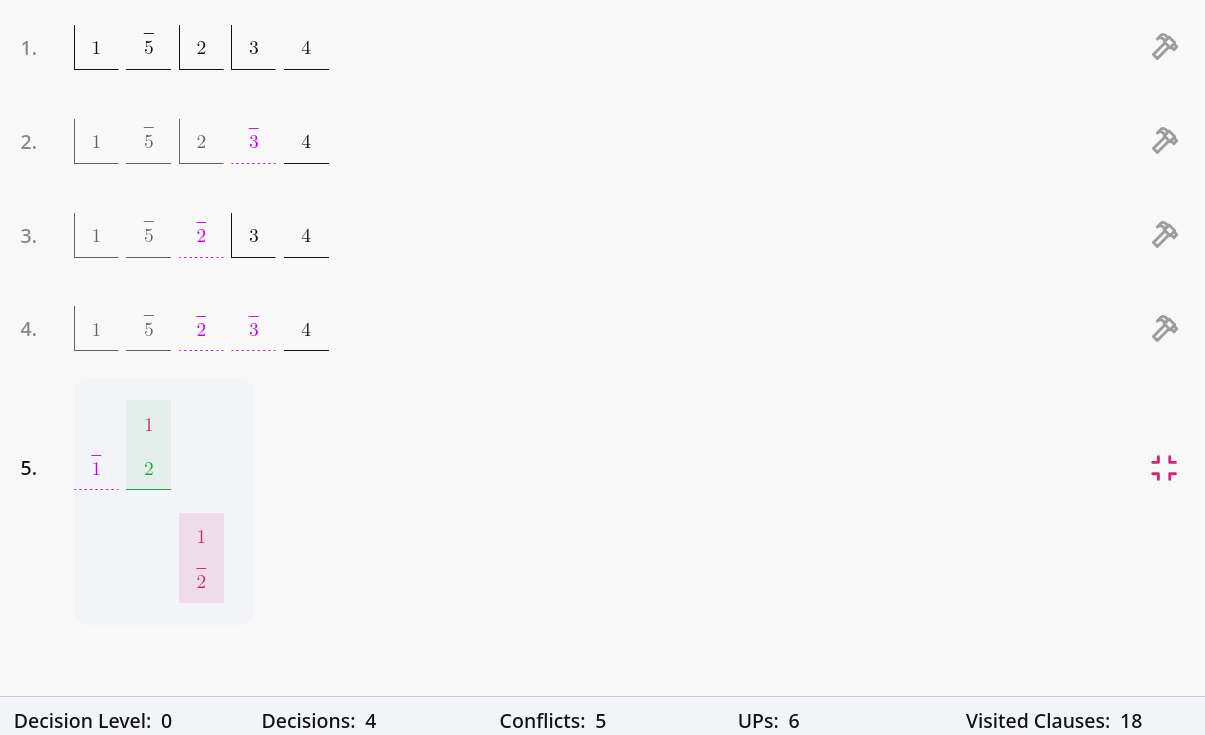}
                \caption{DPLL}
                \label{fig:dpll-b}
            \end{subfigure}
        \end{minipage}
    \end{minipage}

    \vspace{0.3cm}
    \caption{Comparison of search processes over the formula $F$. Figures (a) and (b) represent the search space of the Backtracking and DPLL algorithms.}
\end{figure}

\begin{figure}[t]
    \centering
    \begin{minipage}{\linewidth}
        \centering
        
        \begin{minipage}{0.65\linewidth}
            \centering
            \begin{subfigure}{\linewidth}
                \centering
                \includegraphics[width=0.9\linewidth]{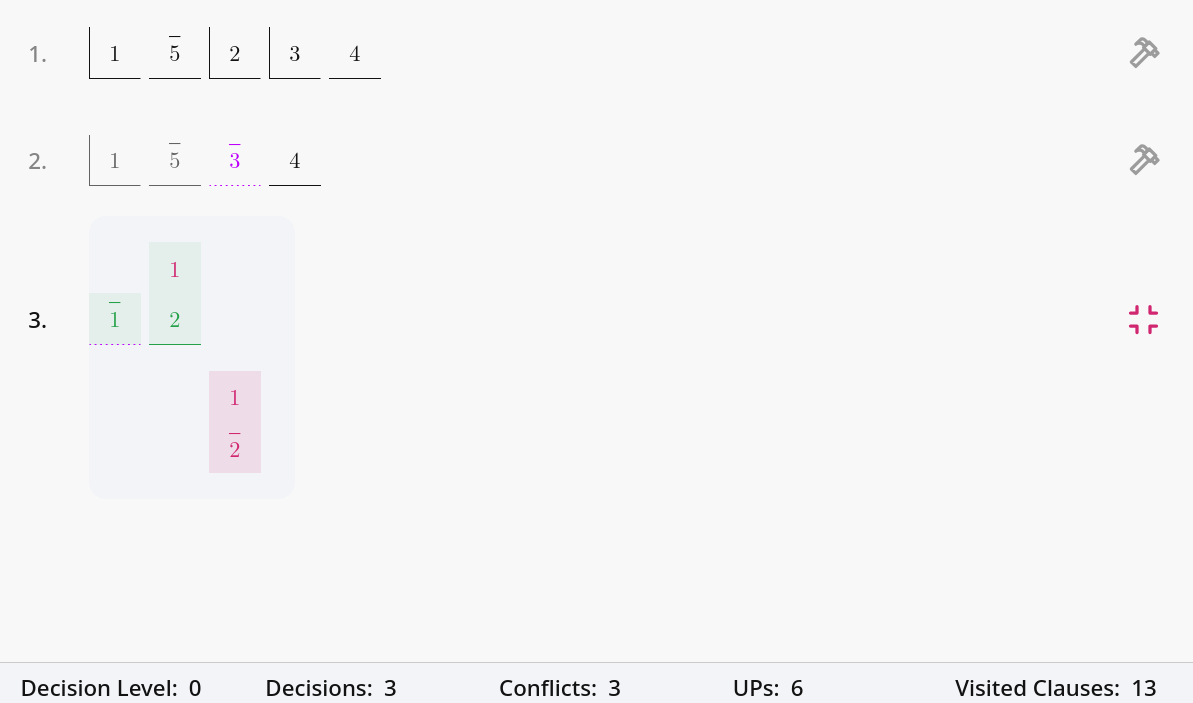}
                \caption{CDCL search space.}
                \label{fig:dpll-c}
            \end{subfigure}

            \vspace{0.7cm} 

            \begin{subfigure}{\linewidth}
                \centering
                \includegraphics[width=0.9\linewidth]{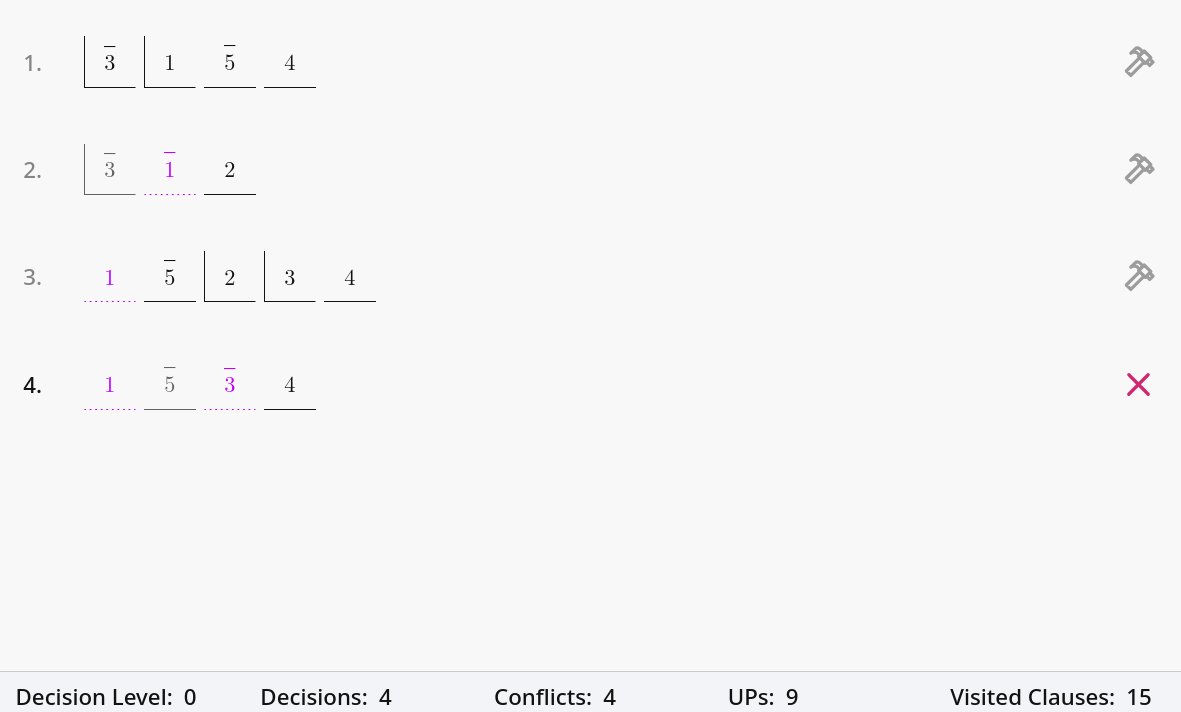}
                \caption{CDCL (starting with $\overline{3}$).}
                \label{fig:dpll-e}
            \end{subfigure}
        \end{minipage}
        \hfill
        \begin{minipage}{0.33\linewidth}
            \centering
            \begin{subfigure}{\linewidth}
                \centering
                \includegraphics[width=\linewidth]{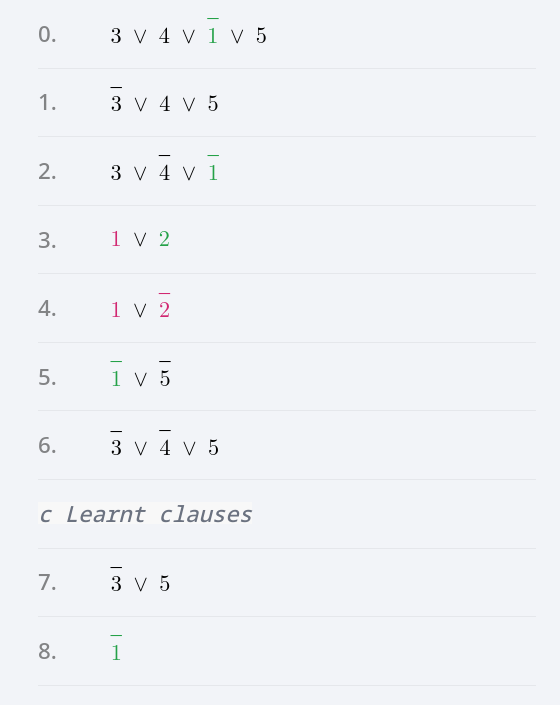}
                \caption{CNF instance.}
                \label{fig:dpll-d}
            \end{subfigure}
        \end{minipage}

    \end{minipage}

    \vspace{0.3cm}
    \caption{Comparison of search processes over the formula $F$. Figures (a) and (b) represent the search space under different initial decisions, while Figure (c) displays the updated CNF due to the visited search space presented in Figure~\ref{fig:dpll-c}.}
\end{figure}

The formula $F$ contains five variables and seven clauses. Using simple backtracking, it takes sixteen conflicts to be aware that the formula is unsatisfiable (see Figure~\ref{fig:dpll-a}).
Once the previous concepts are kept clear, we will proceed with the explanation of DPLL algorithm and its visual representation in SAT-IT employing a comparison with backtracking. In here, we intend to present in detail the occurrence lists, as it is the visual method that captures the detection of unit clauses and clauses that became violated by the current assignment.  DPLL required five conflicts to demonstrate the unsatisfiability of the formula $F$, the reason for this reduction is UP, which entails many of the literals as visiting the occurrence of the complementary of the latest assignments (see Figure~\ref{fig:dpll-b}). Next, we will detail how learning from conflicts leads to a more efficient way of finding a solution or demonstrating unsatisfiability. The execution of CDCL over the formula $F$ is shown in Figure~\ref{fig:dpll-c}. In this example, the assignments for trails one and two lead to two conflictive clauses. Due to these conflictive clauses, in order, conflict analysis was initiated, which resulted in the derivation of clauses $(\overline{3} \lor 5)$ and $(\, \overline{1}\,)$ through resolution. These two clauses are learnt, resulting in the update of the formula, i.e., $F' = F \land (\overline{3} \lor 5) \land (\,\overline{1}\,)$ (see Figure~\ref{fig:dpll-d}). From the last trail, we reached a dead end, where no decision can be undone, stating that $F$ is unsatisfiable. Also, in trail two of Figure~\ref{fig:dpll-c}, after the conflict analysis in trail one, not one but two decision levels were undone by backjumping.

Following the CDCL overview, we will show that most of the execution time is spent traversing occurrence lists to visit clauses that can potentially trigger a UP or be falsified. Hence, it is of the upmost importance to optimize this process. To address this, we will present the 2WL scheme for minimizing the number of visited clauses. By watching two literals in every clause, it is possible to ensure that a clause has not become unit or falsified. For the formula $F$, enabling the 2WL scheme in CDCL avoids visiting five less clauses than basic CDCL. For such a small instance, this is already a notable boost in performance. The effect of enabling 2WL is appreciated by the statistics that appear in the bottom part of the application.

Furthermore,  we will also demonstrate that even with the enhancements of 2WL in CDCL, certain structures like the Pigeonhole Principle became difficult to solve. The hardness of solving such problems is inherent in the proof system under which SAT-solvers are based (resolution). Such problems are difficult because counting for SAT is one of its limitations.

We consider that \emph{what-if} scenarios are one of the strengths of SAT-IT. A \emph{what-if} scenario allows the application users to interact with the solving procedure by enabling manual assignments instead of the default assignment that would be done by the implemented heuristic. The assignment of one literal instead of another can lead to a totally different search space. For example, by using plain CDCL and the formula $F$, if the first literal decided is $\overline{3}$---instead of the literal $1$ given by the automatic assignment heuristic---as shown in Figure~\ref{fig:dpll-e}, a different search space would have been traversed to show the unsatisfiability of the formula.

Alongside the functionalities and the flow management of SAT-IT already presented, the tool presents other utilities that may be worth mentioning in the presentation, such as literal level breakpoints, statistics preview, and more. As SAT-IT is in continuous development, other interesting features may appear in the talk; for example, the introduction of a view for displaying the implication graph for a better understanding of what led to a conflict in CDCL.



\bibliography{lipics-v2021-sample-article}

\appendix

\newpage

\section{Online Resources}
\label{app:online-resources}
The online resources of the project can be found on:
\begin{itemize}
    \item \href{https://github.com/udg-lai/sat-it}{Source Code}
    \item \href{https://udg-lai.github.io/sat-it/}{SAT-IT Website}
\end{itemize}

\section{Architecture}
\label{app:architecture}

\subsection{State Machine}

\begin{figure}[h]
\centering
\scalebox{0.8}{
    \centering
    \includegraphics[width=0.5\linewidth]{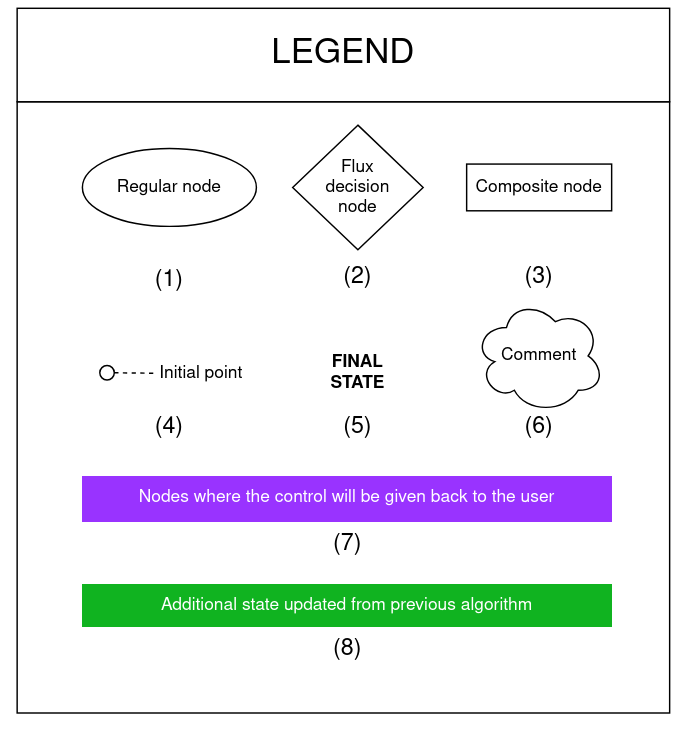}
    }
    \caption{Legend of the SAT-IT' State Machines.}
    \label{fig:legend}
\end{figure}

The deterministic nature of all SAT-solving algorithms implemented in SAT-IT allows them to be modelled as state machines. Moreover, the requirement for user interaction to control their execution flow motivates to represent these algorithms as state machines rather than as standard procedures. This design choice facilitates a smooth progression from Backtracking to DPLL and ultimately to CDCL and its extension with two watch literals, while preserving the interactive nature of the system.

\begin{landscape}
\begin{figure}[tpbh]

    \centering
    \includegraphics[height=0.95\textheight]{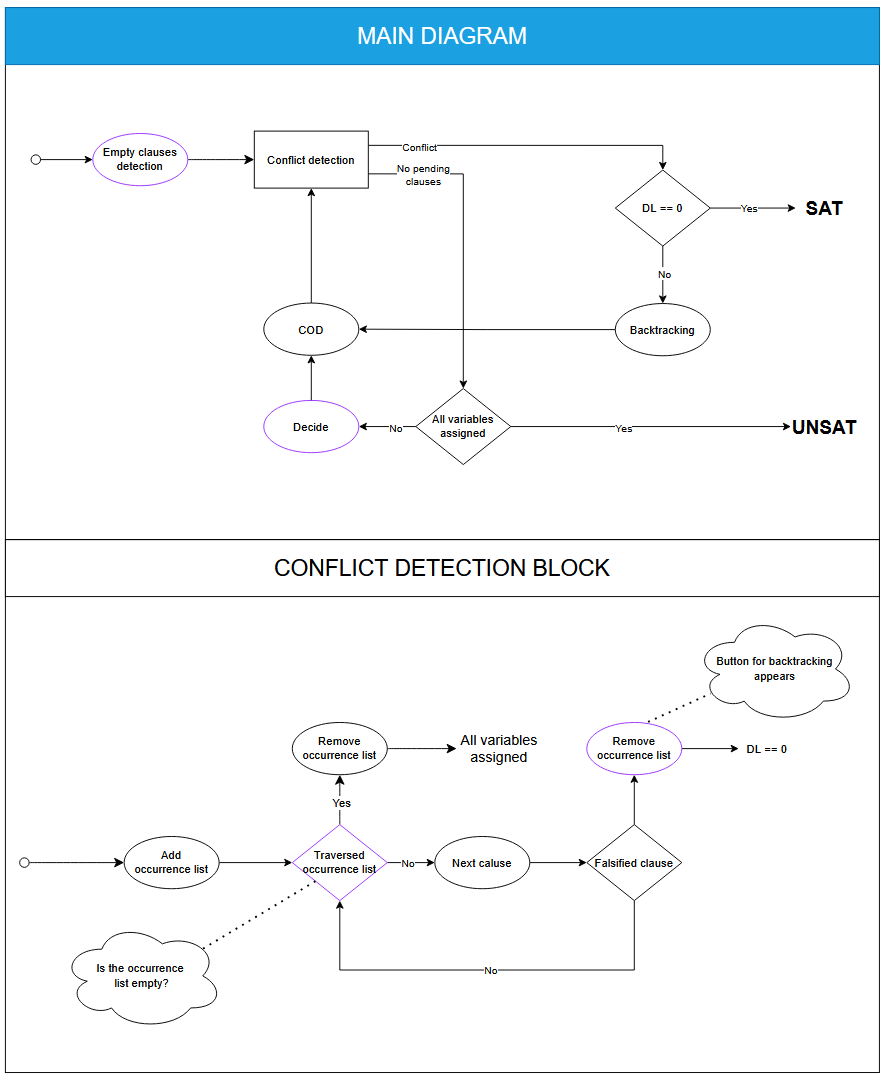}
    \caption{Backtracking state machine diagram}
    \label{fig:backtracking-sm}
\end{figure}
\end{landscape}

\begin{landscape}
\begin{figure}[tpbh]

    \centering
    \includegraphics[height=0.95\textheight]{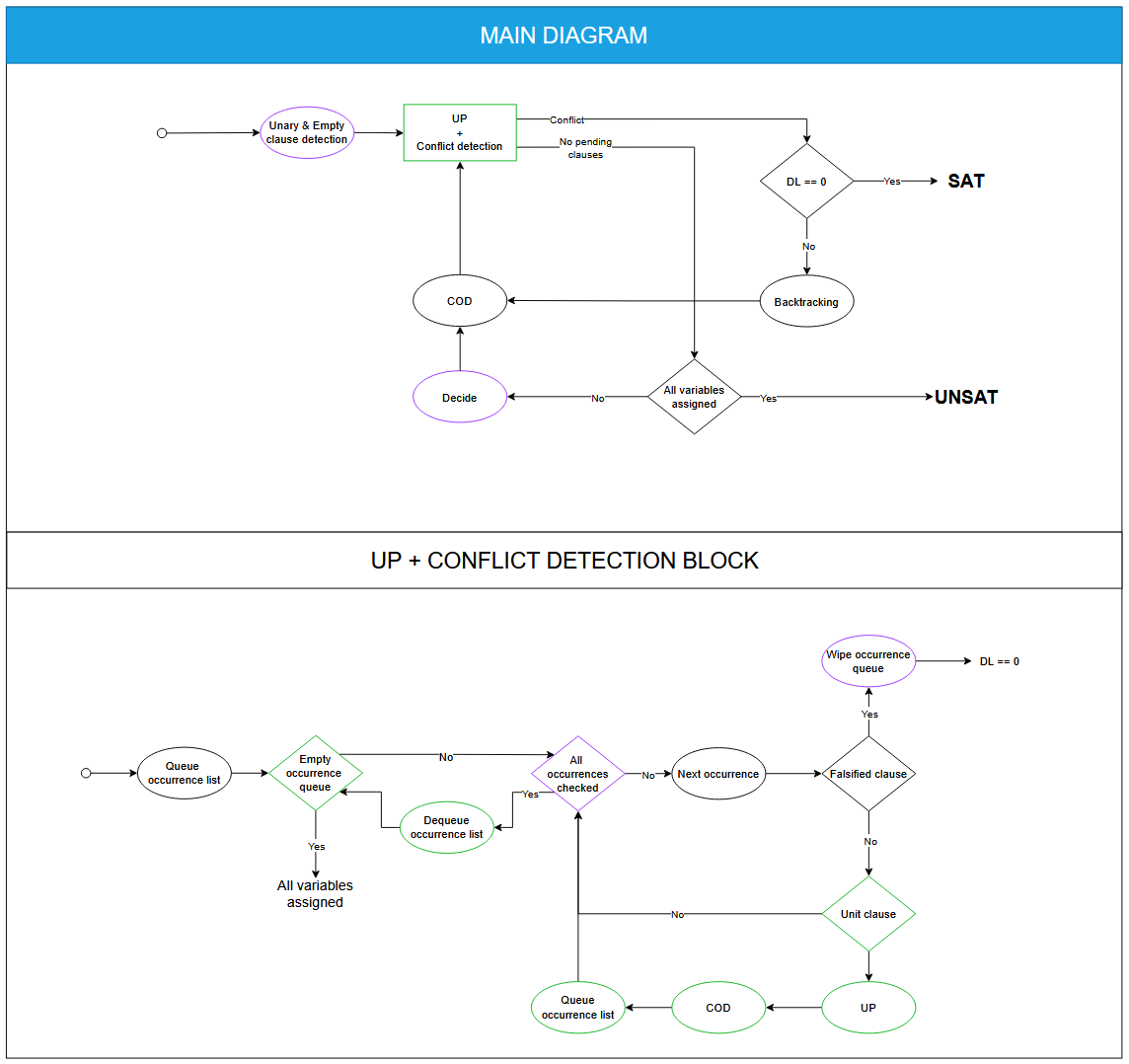}
    \caption{DPLL state machine diagram}
    \label{fig:dpll-sm}

\end{figure}
\end{landscape}

\begin{landscape}
\begin{figure}[tpbh]

    \centering
    \includegraphics[width=1.0\linewidth]{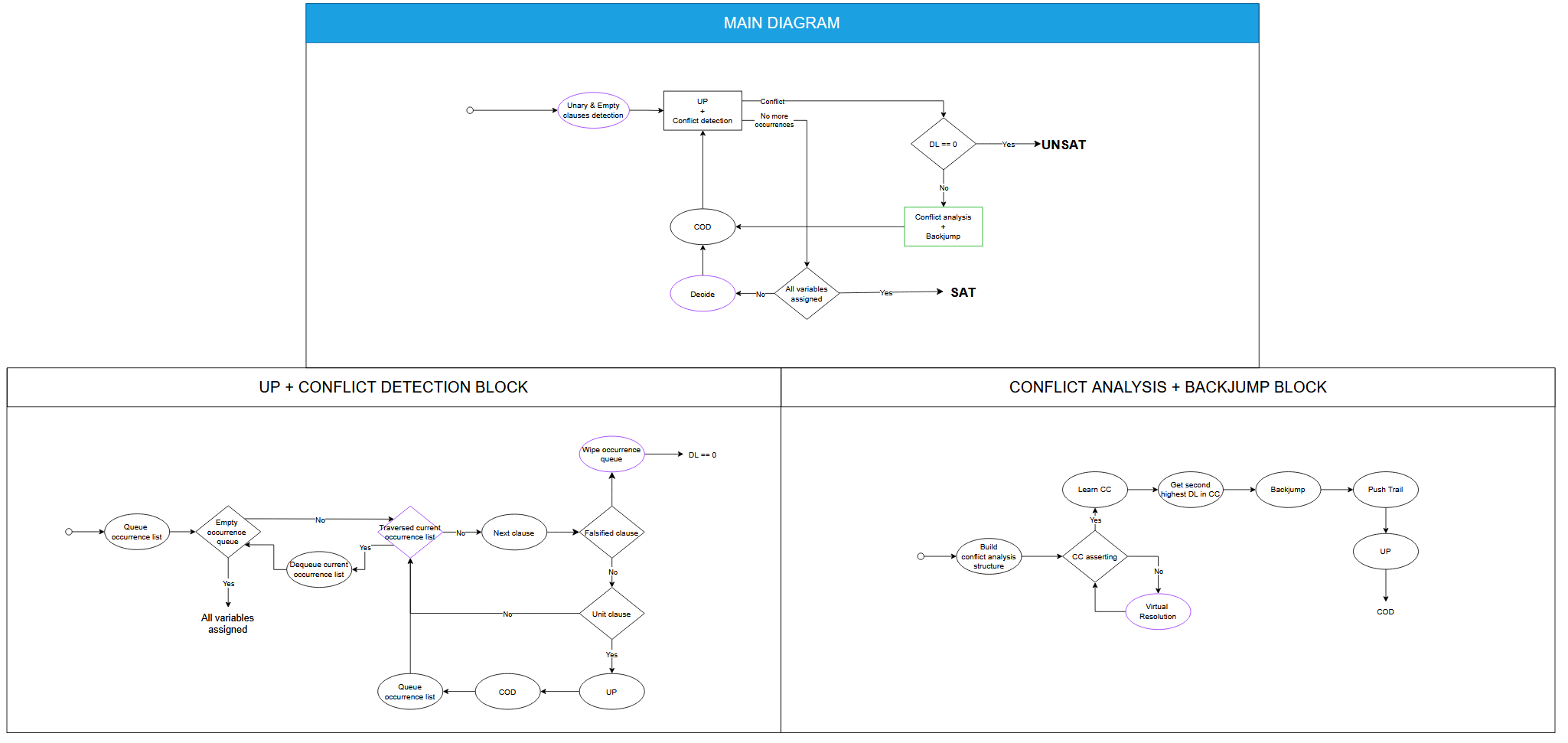}
    \caption{CDCL state machine diagram}
    \label{fig:CDCL-sm}

\end{figure}
\end{landscape}

\begin{landscape}

\begin{figure}[tpbh]

    \centering
    \vspace{1.75cm}
    \includegraphics[width=1.0\linewidth]{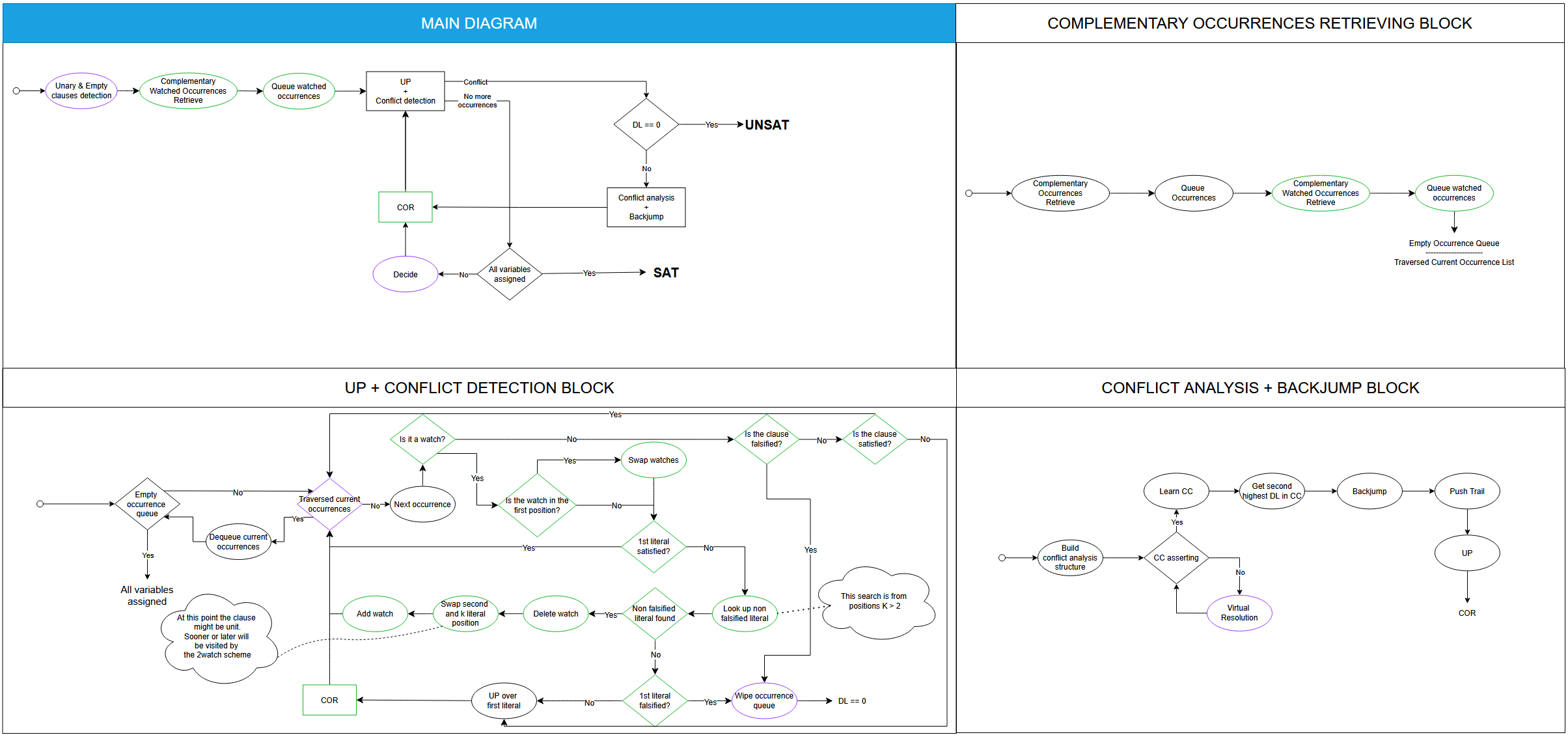}
    \caption{CDCL + 2-Watch Literals state machine diagram}
    \label{fig:CDCL2W-sm}

\end{figure}
\end{landscape}

Figures ~\ref{fig:backtracking-sm},~\ref{fig:dpll-sm},~\ref{fig:CDCL-sm} and~\ref{fig:CDCL2W-sm} illustrate the state machine of the solving algorithms implemented in SAT-IT. To facilitate a clearer understanding, we explain the conceptual components that may appear in the state diagrams using the legend illustrated on Figure~\ref{fig:legend}.

\begin{description}
    \item[1) Regular node:] An atomic unit within a solving procedure.
    \item[2) Flux decision node:] A branching point in the state machine's execution flow.
    \item[3) Composite state:] A high-level abstraction of the solving process that encapsulates multiple regular and decision nodes.
    \item[4) Initial point:] The entry node marking the beginning of the state machine execution block.
    \item[5) Final state:] The termination point of the process, reached once a model is identified or the problem is proven unsatisfiable.
    \item[6) Comment:] An annotation used to provide conceptual clarification for specific components.
    \item[7) Violet border nodes:] Specialized regular or decision nodes that interrupt the standard execution flow to return control to the user.
    \item[8) Green border nodes:] Regular or decision nodes newly introduced by the algorithm. For instance, the CDCL algorithm introduces the composite node that contains the conflict analysis and backjumping that DPLL does not have.
\end{description}

\subsection{Solver Machine}

The solver machine acts as a controller between the user-interaction through the application and the solving procedures. This design allows each solver to define and manage additional parameters or logic unique to its solving algorithm, supporting a modular and extensible architecture.

The primary concepts that any solver machine grasps are:

\begin{description}
    \item[\texttt{transition}:] Moves from one state to another of the state machine.

    \item[\texttt{step}:] Encapsulates the particularity of visiting the state machine of each solver. This is a logic step and typically makes multiple transitions over its state machine.

    \item[\texttt{step-by-step}:] Implemented as higher-order-function. Encapsulates the execution of many logic steps until a certain halt condition is accomplished.

    \item[\texttt{transition-by-event}:] Encapsulates the particularity of visiting the state machine of each solver by running a single step or many of them depending on the event sent from the application because of a user interaction.
\end{description}

\newpage

\end{document}